\documentclass[10pt,article]{article}
\usepackage[top=0.85in,left=0.75in,footskip=0.75in,marginparwidth=2in]{geometry}
\usepackage{physics}
\usepackage{nameref,hyperref}
\usepackage{graphicx}
\usepackage{color}
\usepackage{abstract,lipsum}
\usepackage{cite}
\usepackage{amsmath}
\newcommand{\angstrom}{\textup{\AA}}

\usepackage[aboveskip=1pt,labelfont=bf,labelsep=period,singlelinecheck=off]{caption}

\makeatletter
\renewcommand{\@biblabel}[1]{\quad#1.}
\makeatother

\usepackage{lastpage,fancyhdr,graphicx}
\usepackage{epstopdf}
\usepackage{wrapfig}
\usepackage[pscoord]{eso-pic}
\usepackage[fulladjust]{marginnote}
\reversemarginpar

\begin{document}
\vspace*{0.35in}

\begin{flushleft}
{\LARGE
\textbf\newline{ Phase-controlled photon drag in a slow-light moving medium}
}
\newline
\\
Seyedeh Hamideh Kazemi\textsuperscript{1},
Mohammad Mahmoudi\textsuperscript{1,*}
\\
\bigskip
\textsf{1} Department of Physics, University of Zanjan, University Blvd., 45371-38791, Zanjan, Iran
\\
* mahmoudi@znu.ac.ir

\end{flushleft}
\begin{abstract}
In recent years, photon drag has attracted enormous attention owing to both fundamental and practical interests. In this paper, by presenting a density-matrix approach, we have theoretically demonstrated an enhanced photon drag in a moving atomic medium. By incorporating of the interference of spontaneous emission, properties of the medium can be easily controlled by the relative phase of applied fields so that a large group index along with a transparency or even a gain can be achieved. As photon drag is proportional to group index, the enhancement can be also found in the dragging effect. Applications of the enhanced dragging effect can be found for efficient modulators of light, position control, and detection of slow motion.
\end{abstract}
\vspace*{0.2in}

\section{Introduction}

The phase velocity of light traveling in a moving medium with velocity $\mathrm{v}$ deviates from the velocity of light in vacuum ($c$). Naively, one would then expect that the phase velocity of light, $\mathrm{v}_{p}$, can be formulated by the Newtonian velocity addition $\mathrm{v}_p= c/n +\mathrm{v}$, in the case of the low-speed limit $\mathrm{v} \ll c$, with $n$ being as the index of refraction of the moving medium. Today, thanks to works of pioneers, we know that once light propagates through a moving medium, it can be dragged in either transverse or longitudinal direction. In fact, the deviation from the Newtonian velocity addition was predicted by Fresnel in 1818 on the basis of an elastic aether theory; Based on Young's description of the wave character of light \cite{young}- with its consequence that the speed of light is inversely proportional to the index of refraction- and his suggestion that this index may depend on the concentration of aether within the medium, Fresnel \cite{fresnel} concluded that light traveling in a moving medium would be dragged so as to have an additional component of velocity $\mathrm{v} \,(1-n^{-2})$ in the direction of the medium. Fresnel's prediction for the longitudinal case was experimentally verified by Fizeau \cite{fiz} in a flowing water tube experiment. Over 100 years later, Jones \cite{jones} presented an experimental demonstration of lateral photon drag for the case of a light beam transmitting near the edge of a spinning glass disk with transverse velocity $\mathrm{v}$, using the formula $\Delta x= (n_g-n_{r}^{-1}) \, \mathrm{v} L /c $, where $n_r$ and $n_g$ are, respectively, the phase and group refractive indexes and $L$ is the length of the medium. A mechanical rotation of the medium is subsequently predicted to induce a rotary photon drag, which was verified by Jones \cite{jones2}. He observed rotation of linear polarization state of light through a small angle given as $\theta_r= (n_g-n_{r}^{-1}) \, \omega_s L / c $, where $\omega_s$ is the spin angular velocity of the dielectric medium. Here, the rotation of a linear polarization state originates from a phase difference between the right- and left-handed circular polarization states, which corresponds to the spin angular momentum states of light \cite{6}. However, light can also carry orbital angular momentum arising from helical phase fronts and the associated azimuthal component of the Poynting vector \cite{7} which account for the rotation of the transmitted image \cite{6,8}.

The phenomenon of light dragging by a moving host medium, sometimes called Fresnel drag, can be explained by the special theory of relativity. As light enters a medium, its phase velocity with respect to the reference frame attached to the medium changes to $c/n$. Thus, the phase velocity of light with respect to the stationary laboratory frame is given by the relativistic addition of two velocities $c/n$ and $\mathrm{v}$, and, therefore, can be written as $\mathrm{v}_p = (c/n\pm \mathrm{v}) / (1+\mathrm{v}/nc) \simeq c/n \pm F \mathrm{v}$ with $F=1-n^{-2}$ being as the Fresnel drag coefficient. In a dispersive medium, the Fresnel drag coefficient has to be modified. Because of the Doppler effect, the frequency of light $\nu$ measuring in the laboratory frame becomes $\nu^{'} \simeq \nu \,(1\mp \mathrm{v}/c)$ and the refractive index has to be that appropriate to the modified frequency. Therefore, the drag coefficient is given by $\mathrm{v} \,(1-n^{-2} + (\nu / n)\, dn/d\nu )$ which was first pointed out by Lorentz \cite{lorentz}. After that, Zeeman and his collaborators \cite{zeeman} performed a series of experiments over a period of more than 10 years in order to measure light drag accurately. They observed the predicted contribution of dispersion on the light-drag effect by moving a $1.2$-m-long glass rod at speed 10m/s. Noting that the effect of dispersion in normal materials with low dispersive properties is so small that the magnitude of this contribution can be disputed \cite{Lerche}.

In the intervening years, the Fresnel drag has attracted enormous attention because of its fundamental and practical interest, for instance, improving the measurement’s accuracy \cite{12,13}, differentiating it from competing effects such as the Sagnac effect \cite{14,15}, dragging massive particles such as neutrons \cite{16}, and proposing dielectric analogs of gravitational effects \cite{17}. In a non-dispersive medium like water or glass, the drag coefficient is only on the order of one and, therefore, a few meters long tube was used in early Fizeau’s water tube experiment \cite{fiz} in order to have an observable effect. Further experiments used a spinning glass rod in a ring resonator in order to improve the detection sensitivity \cite{12,66}. Refrences \cite{6,8,63,65} demonstrated photon drag by experiments in slow-light ruby media. It is imperative to point out a remarkable work of Arnold \textit{et al.} \cite{6} in which a considerable enhancement in rotary photon drag was observed in a slow-light spinning ruby medium. 

Light pulses propagating in atomic media with a highly dispersive nature, including alkali atomic vapors, can also experience a dragging effect. However, the photon drag can be enhanced in dispersive atomic vapours, a large dispersion is usually accompanied by strong absorption of light at the resonance frequency. A recent experiment shows the phase velocity dragging in a hot rubidium (Rb) vapour by shifting the frequency away from the resonance to avoid the absorption and improves the dragging coefficient by two orders of magnitude \cite{safari}. Their results are in good agreement with the theoretical prediction indicating that the strength of the light drag effect is proportional to the group index of the moving medium in the limit of large group indexes. At about the same time, Kuan \textit{et al.} \cite{72} observed a further improvement in phase velocity dragging using a three-level cold atomic medium with the scheme of the electromagnetically induced transparency (EIT). In 2017, Khan \textit{et al.} \cite{khan} investigated the rotary photon drag and surface plasmon polaritons drag in a four-level N-type atomic medium and showed that control of these drag coefficients has been accomplished either by changing the spinning velocity of the host medium or by changing the control field's Rabi frequency. More recently, Iqbal \textit{et al.} \cite{iqbal} presented a theoretical demonstration of enhanced photon drag in a slow-light moving medium; the incorporation of Kerr-type nonlinearity to the atomic medium endows its absorption spectrum with the feature of EIT amplification, resulting in a significant enhancement in the rotary photon drag. It has been shown that the dragging effect can be controlled by the parameters of the driving beams, such as the Rabi frequency and the detuning. In this context, there are proposed experiments using the dispersive media for studying the motional sensing, transverse light dragging and laboratory analog of astronomical systems, such as event horizon in the black hole \cite{17,zimmer,sensing,rocca,horiz}. In spite of the pronounced success, there still exists a continuing need for enhancement in photon drag. 
%It has been shown that the light-drag enhancement is due to the large group index of a Rb vapor and can be understood by means of the Doppler effect. 

Coherent or incoherent interaction between electromagnetic waves and atoms induces coherence among atomic states, leading to interesting quantum interference effects. In this context, the relative phase of applied fields in an atomic system is an important parameter for controlling the atomic coherence. It is well known that the optical properties of a closed-loop atomic system interacting with laser fields are completely phase-dependent \cite{7m,8m,9m,kazemi1}. The phase-dependent behaviour can be also induced by quantum interference due to spontaneous emission of an atom with two closely lying levels \cite{4d,1d,2d,5d,11d,6d,9d,8d,12d,7d,10d,sahrai,dutta1,dutta2,Zohravi}. Such coherence can be created by the interference of spontaneous emission, called spontaneously generated coherence (SGC), of either a single excited level to two close-lying atomic levels \cite{1d} or two close-lying atomic levels to a common atomic level \cite{2d}. In ladder-type system, this type of coherence can be created in a nearly-spaced atomic levels case \cite{4d}. 

On the other hand, in the past three decades, controlling group velocity of light has attracted a lot of interest owing to its potential applications, such as tunable optical buffers, optical memory and enhancing the nonlinear effect. Up to now, numerous experimental and theoretical works have been devoted to control it in materials such as atomic medium \cite{4m,8m,Steinberg,Kash,Sahrai4,9m}, optomechanical system \cite{opto,opto2}, and Landau-quantized graphene \cite{grap1,graph2}. Slow light can be used in telecommunication applications such as controllable optical delay lines, optical buffers, true time delay methods for synthetic aperture radar, development of spectrometers with enhanced spectral resolution and optical memories. Moreover, a great deal of experiments confirmed that it is possible for optical or electrical wave pulses to travel through materials with group velocities greater than $c$. Another interesting scenario in light propagation concerns the situation which the group velocity of light can even become negative. It is worth mentioning that the superluminal light propagation does not violate Einstein’s theory of special relativity since the energy and information flow do not exceed $c$ \cite{Stenner}. An ideal condition for practical light propagation is a region in which the light pulse should not attenuate or amplify, primarily due to fact that pulse propagation does not possible in the presence of a large absorption.  

As mentioned above, an considerable enhancement of the photon drag in highly-dispersive media has been observed in experimental and theoretical works. One of the main challenges is being able to easily control this dragging effect, as previous studies focused on controlling methods of the photon drag via frequency and intensity of applied fields. In this paper, we exploit the fact that the properties of the medium and hence the group velocity can be dramatically modified by the relative phase of applied fields, allowing us to simultaneously control and further enhance the photon drag. Additionally, the dispersion and absorption of a weak probe field in the system can be dramatically affected by the relative phase so that the enhanced photon drag along with a gain can be achieved for a proper choice of the relative phase.

\begin{figure*}
\centering
\includegraphics[width=11cm]{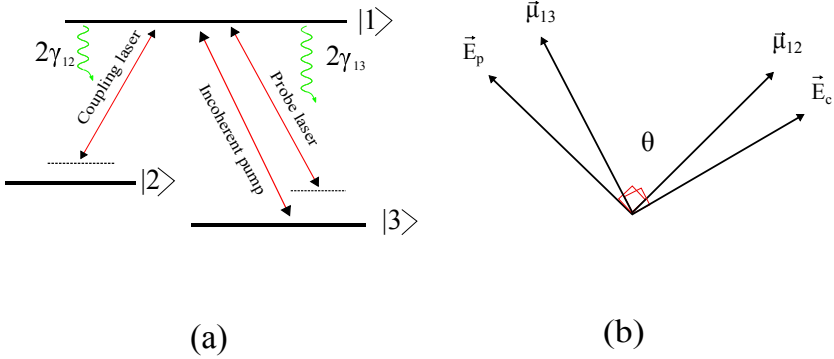}
\caption{Schematic diagram of a three-level $\Lambda$-type atomic system. The probe and coupling fields have been applied between levels $\vert 1\rangle \leftrightarrow \vert 3\rangle$ and $\vert 1\rangle \leftrightarrow \vert 2\rangle$, respectively. The parameters $2 \gamma_{13}$ and $2 \gamma_{12}$ denote the spontaneous decay width from the respective levels. (b) The arrangement of field polarization required for a single field driving one transition if dipoles are non-orthogonal. Here, $ \vec{E}_{c(p)}$ is the amplitude of the coupling (probe) fields and $\theta$ represents the angle between two dipole transition moments. }
\label{figure1}
\end{figure*}

\section{The model}

We consider a three-level $\Lambda$-type atomic system consisting of two nearly degenerate levels $\vert 2 \rangle$ and $\vert 3 \rangle$, and an excited level $\vert 1 \rangle$ [see Fig.~\ref{figure1}(a)]. The coupling field with a frequency $\omega_{c}$ and the probe field with frequency $\omega_p$ are applied on the transition $\vert 2\rangle \leftrightarrow  \vert 1\rangle$ and $\vert 1\rangle \leftrightarrow  \vert 3\rangle$, respectively. Moreover, the transition $\vert 3\rangle \leftrightarrow  \vert 1\rangle $ is pumped with a rate $2 \Lambda$ by a bidirectional incoherent pumping. The spontaneous decay width from the level $\vert 1 \rangle$ to $\vert 2 \rangle$ and $\vert 3 \rangle$ are denoted by $2 \gamma_{12}$ and $2 \gamma_{13}$, respectively. Here, $\theta$ represents the angle between two dipole transition moments. As the existence of the SGC effect depends on non-orthogonality of the dipoles $\vec{\mu}_{13}$ and $\vec{\mu}_{12}$, we have to consider a scheme where each field acts only on one transition (Fig.~\ref{figure1}(b)). 

Using the von-Neumann equation, we can easily arrive at the density-matrix equation of motion in the rotating-wave and the electric-dipole approximation \cite{dutta1}

\begin{subequations}
\begin{eqnarray}
\dot{\rho}_{11}&=& -2 (\gamma_{12}+\gamma_{13}+\Lambda )\rho_{11}+ 2 \Lambda \rho_{33} +i \Omega_{c} \rho_{21}- i \Omega_{c}^{*} \rho_{12} +i \Omega_{p} \rho_{31} -  i \Omega_{p}^{*} \rho_{13}, \\
\dot{\rho}_{22} &=&  2 \gamma_{12} \rho_{11} +i \Omega_{c}^{*} \rho_{12} -i\Omega_{c} \rho_{21},\\
\dot{\rho}_{33} &=&   2 (\gamma_{13} + \Lambda ) \rho_{11} -2\Lambda \rho_{33}+i \Omega_{p}^{*} \rho_{13} -i \Omega_{p} \rho_{31}, \\
\dot{\rho}_{12}&=&  -  ( \gamma_{13}+ \gamma_{12}+ \Lambda + i\Delta_{c}) \rho_{12} + i \Omega_{p} \rho_{32} + i  \Omega_{c} (\rho_{22}-\rho_{11}), \\ 
\dot{\rho}_{13}&=&  -  ( \gamma_{13}+ \gamma_{12}+ 2 \Lambda + i\Delta_{p}) \rho_{13} + i \Omega_{c} \rho_{23} + i  \Omega_{p} (\rho_{33}-\rho_{11}), \\ 
\dot{\rho}_{23}&=&  -  [ i(\Delta_{c}- \Delta_{p} )-\Lambda ]  \rho_{23} + i \Omega_{c}^{*} \rho_{13} - i  \Omega_{p} \rho_{21} + P\,  \rho_{11}.
\end{eqnarray}
\label{eq1}
\end{subequations}

The remaining equations follow from the constraints $\rho_{ij}=\rho^{*}_{ji}$, $\sum _{i} \rho_{ii}=1 $ and the parameters $\Delta_{c}=\omega_c-\omega_{12}$ and $\Delta_{p}=\omega_p-\omega_{13}$ denote the detunings between applied fields and  corresponding atomic transitions. Noting that the expression for Rabi frequency is defined as $\Omega_{c(p)}=( \vec{\mu}_{12(13)}.  \vec{E}_{c(p)})/{\hbar} =\Omega_{0c(0p)} \sin \theta $ with $\Omega_{0c(0p)}$ being as Rabi frequency of the coupling (probe) field without the SGC. The term including $ P=2 \eta \,\sqrt{\gamma_{13} \gamma_{12} }$ with $\eta=\eta_{0} \cos \theta$ accounts for the effect of quantum interference due to spontaneous emission. For the geometry shown in Fig.~\ref{figure1}(b), $\theta$ is always nonzero, though it could be small. It should be noted that only for small energy spacing between levels $\vert 2 \rangle$ and $\vert 3 \rangle$, the SGC effect becomes remarkable in such a way that we have $\eta_{0} = 1$ if those levels lie so closely, otherwise $\eta_{0}=0$.  

While a $\Lambda$-type system with well-separated levels do not depend on the relative phase of two applied fields, in case of closely-spaced levels, it becomes quite sensitive to the relative phases due to existence of the SGC term. By considering $\Omega_c=\tilde{\Omega}_c e^{-i \varphi_c}$,  $\Omega_p=\tilde{\Omega}_p e^{-i \varphi_p}$-with $\varphi_c$ and $\varphi_p$ being as the phase of the coupling and probe fields, respectively- and redefining the new atomic density-matrix variables in equations (\ref{eq1}), we obtain equations for redefined density-matrix elements $\tilde{\rho}_{ij}$ which are identical with those equations; only the SGC parameter $\eta_0$ is replaced by real parameters $\eta_{0} e^{i \Delta \varphi}$, $\Omega_c$ and $\Omega_p$ are replaced by $\tilde{\Omega}_c$ and $\tilde{\Omega}_p$, with $\Delta \varphi= \varphi_p- \varphi_c$ being as the relative phase of applied fields.

%We assume fields to have equal frequencies $\omega_1=\omega_2=\omega$ but different phases $\varphi_1=\varphi_2=\varphi$, unless otherwise stated explicitly. 

On the other hand, the response of the atomic system to applied fields is determined by the susceptibility $\chi$, which is defined as \cite{scully}
\begin{equation}
\chi(\Delta_{p}) = \frac{ 2 N \vert \mu_{13}^2 \vert}{ \hbar \epsilon_0 \tilde{\Omega}_p} \, \tilde{\rho}_{13}(\Delta_{p}) = \chi^{'}(\Delta_{p}) + i \chi^{''}(\Delta_{p}),
\end{equation}
where $N$ and $\epsilon_0$ are, respectively, atom number density in the medium and the permittivity of free space. Real and imaginary parts of the susceptibility correspond to the dispersion and absorption, respectively. In order to calculate the susceptibility in a realistic situation, we consider the data of experiments on a medium consisting of cold three-level atoms \cite{hue,Budker,oscar}. By considering parameters $ N \vert \mu_{13}^2 \vert / (4 \hbar \epsilon_0 \gamma_{13}) =0.013$ and $\gamma_{13}=2\pi \times 5$ MHz, we have written relation between the coherence and the susceptibility as $\chi =( 0.1 \, \tilde{\rho}_{13} )/ (\tilde{\Omega}_p /\gamma_{13})$. As is seen, the main observable is the coherence, $\tilde{\rho}_{13}$, which can be obtained by solving the density-matrix equation of motion.

For further discussion, we introduce the group index $n_g=c/\mathrm{v}_g$ and the group velocity $\mathrm{v}_g$ is given as 
\begin{equation}
\mathrm{v}_g=\dfrac{c}{1+ 2 \pi \, \chi^{'} + 2 \pi \omega_p \, \dfrac{\partial}{\partial \omega_p} \chi^{'}}.
\end{equation}
It is clear from this expression that the group velocity of a light pulse can be determined by the slope of the dispersion;  when $\chi^{'}$ is zero and the dispersion is very steep and positive, the group velocity is significantly reduced, leading to the subluminal light propagation. While strong negative dispersion can lead to increase in the group velocity and even to becoming negative, showing the subluminal light propagation. In our notation, negative (positive) values in the imaginary part of susceptibility show the gain (absorption) for the probe field.

\section{Results and discussion}
\begin{figure}
\centering
\includegraphics[width=12cm]{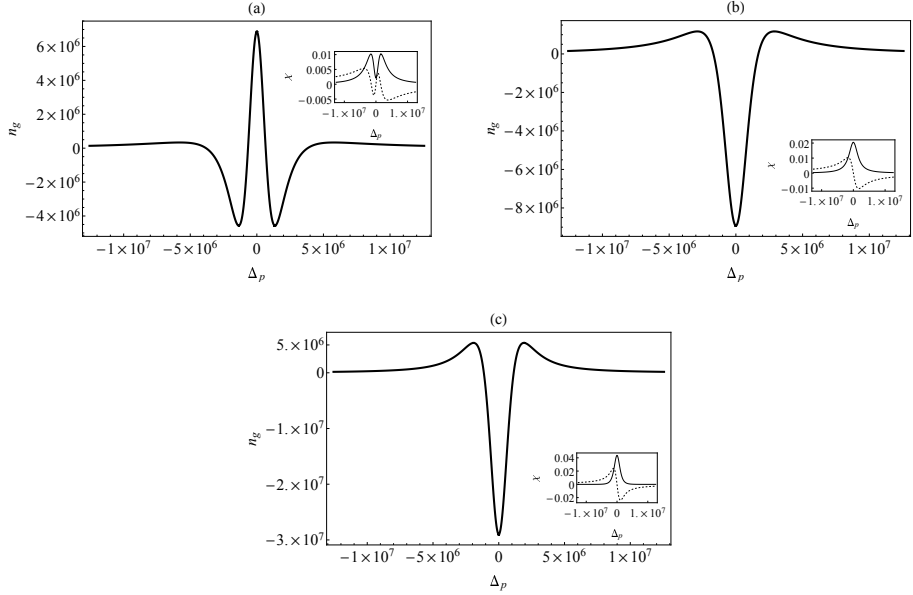}
\caption{Group index versus the probe field detuning $\Delta_p$ for three different SGC parameters: (a) $\eta_0=0.1$, (b) $\eta_0=0.5$ and (c) $\eta_0=0.99$. The selected parameters are $\gamma_{13}=\gamma$, $ \gamma_{12} = 4 \gamma$, $\tilde{\Omega}_p=0.1 \gamma\,\sin \theta \,$, $\tilde{\Omega}_c=3 \gamma \, \sin \theta$, $\Delta_{c}=0$, $\Lambda=0.1 \gamma$ and $\Delta \varphi=0$. }
\label{figure2}
\end{figure}
We now turn towards the steady-state results of the master equations (\ref{eq1}). As the present  model contains a number of parameters that affect the optical properties of the system, and subsequently the photon drag, we first investigate the influence of the quantum interference on the behavior of the system and hence, we shall restrict the range of variation of the parameters involved. In Figs.~\ref{figure2}(b)-\ref{figure2}(c), we plot the group index $n_g$ as a function of the probe detuning $\Delta_p$ for three different values for strength of the interference in spontaneous emission, i.e., the SGC parameters: (a) $\eta_0=0.1$, (b) $\eta_0=0.5$ and (c) $\eta_0=0.99$. In the insets of these figures, we have shown the corresponding profiles for real and imaginary parts of susceptibility to explain the behavior of group index with probe field detuning. The other parameters are $\gamma_{13}=\gamma=2\pi \times 5$ MHz, $ \gamma_{12}= 4 \gamma$, $\tilde{\Omega}_p=0.1 \gamma \,\sin \theta $, $\tilde{\Omega}_c=3 \gamma \, \sin \theta $, $\Delta_{c}=0$, $\Lambda=0.1 \gamma$ and $\Delta \varphi=0$. Here, we have chosen a modest value for the angle between dipole transition moments $\theta=\pi/3$.
%and the frequency of the probe field, $\omega_p=10^{8}$Hz \cite{oscar}. 
\begin{figure}
\centering
\includegraphics[width=12cm]{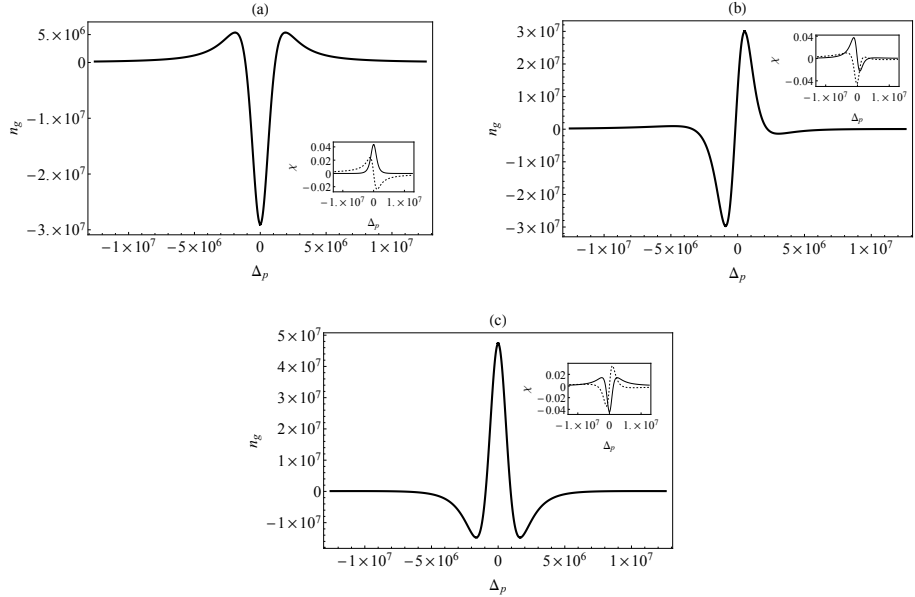}
\caption{ Group index versus the probe field detuning $\Delta_p$ for (a) $\Delta \varphi =0 $, (b) $\Delta \varphi =\pi/2 $, and (c) $\Delta \varphi =\pi $. Inset in each figure shows the imaginary (solid line) and real (dotted line) parts of the corresponding susceptibility. Other parameters are the same as in Fig.~\ref{figure2}(c). }
\label{figure3}
\end{figure} 

Investigation of Fig.~\ref{figure2} shows that in the presence of a small quantum-interference, we get subluminal light propagation around the resonance, i.e. $\Delta_p=0$. The inset shows the variation of  $\chi^{''}=\Im[\chi(\Delta_p)]$ (solid curve) and $\chi^{'}=\Re[\chi(\Delta_p)]$ (dotted curve). The solid curve in the inset shows a partial EIT, as the absorption profile shows a dip around the resonance with a very small absorption at the resonance. The slope of the dotted curve (i.e., dispersion curve) is positive around the resonance, indicating the loss-less subluminal propagation of light. By increasing the interference effect, large variation of $n_g$ occurs around the resonance and we get superluminal light propagation (see Fig.~\ref{figure2}(b) and (c)). The inset of these figures shows the corresponding susceptibility as a function of the probe detuning. For these cases, one can get maximum absorption at the resonance with decreasing magnitude away from the resonance, known as electromagnetically induced absorption (EIA). The dotted line displays that $\chi^{'}$ at the resonance is zero and its slope around the resonance is negative, which implies that the probe laser propagates at superluminal group velocity. Since the photon drag is proportional to the group index, and it is expected to be increased in the former case, we have assumed $\eta_0 =0.99$ in the following of the discussion. 
%{dutta1,dutta2}.

Next, we will demonstrate the crucial role of the relative phase of applied fields in determination of light propagation; Indeed, following figures demonstrate that how group indexes, without changing parameters of fields, can be controlled via the relative phase. As mentioned before, the photon drag has a contribution that is proportional to the group index and, therefore, the photon drag is expected to be controlled via the relative phase. Figure~\ref{figure3} shows the group index versus the probe field detuning with maximum interference ,$\eta_0=0.99$, and for different values of the relative phase of applied fields: (a) $\Delta \varphi =0 $, (b) $\Delta \varphi =\pi/2 $, and (c) $\Delta \varphi =\pi $. In Fig.~\ref{figure3}(a), in spite of  a negative large group index, we observe a large absorption at the resonance and hence, the medium is not suitable for application. As seen in Fig.~\ref{figure3}(b), by tuning the relative phase from 0 to $\pi/2$, the sign of the group index in a small frequency region around $\Delta_p=0$ changes from negative (superluminal) to positive (subluminal). With the change of the relative phase from  $\pi/2$ to $ \pi$, the response of the system changes significantly: We get a gain-assisted subluminal light propagation around the resonance; As the solid and dotted curves in the inset of Fig.~\ref{figure3}(c) represents, around the resonance, a gain is obtained and the slope of the dispersion curve becomes positive and steep. Noting that slope of the dispersion curve away from the resonance would be negative.
\begin{figure}
\centering
\includegraphics[width=12cm]{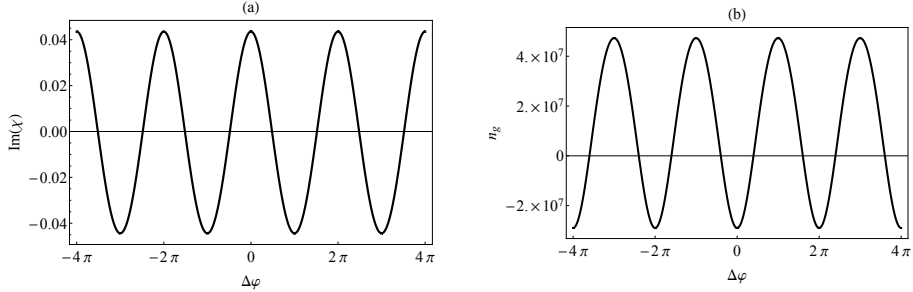}
\caption{ (a) The probe absorption and (b) its group index at the resonance, $\Delta p=0$, as a function of the relative phase of applied fields ($\Delta \varphi$). The other parameters are kept the same as in Fig.~\ref{figure2}(c).}
\label{figure4}
\end{figure} 

We then proceed to provide deeper insight into the influence of the relative phase on the absorption spectrum and group index. To emphasize such effects, we depict in Fig.~\ref{figure4} the probe absorption and the group index at the resonance, when the relative phase of applied fields, $\Delta \varphi$, is varied. It can be seen that both the absorption and the group index are periodical functions of the relative phase, thus by changing it, the probe laser may experience gain, transparency, or absorption. The corresponding variation of the group index with the relative phase shows that it oscillates between negative and positive values. Therefore the relative phase plays an important role in the behavior of the system allowing the atom to exhibit gain-assisted subluminal (superluminal) light propagation. It is worth noting that there are small regions close to $\pm \pi, \pm 3 \pi, ...$ where both the gain and the group index take simultaneously large values.
\begin{figure}
\centering
\includegraphics[width=12cm]{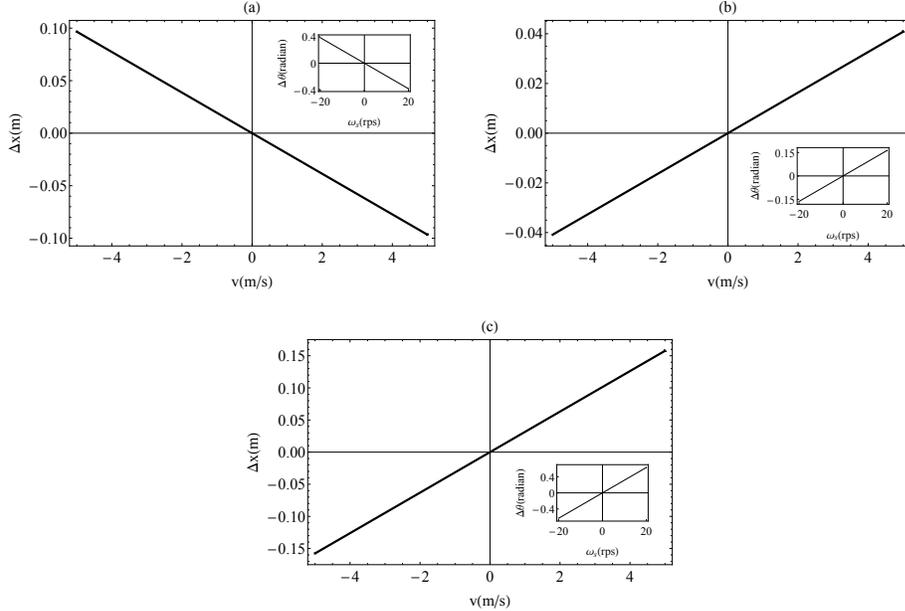}
\caption{Lateral photon drag vs medium transverse velocity $\mathrm{v}$ for $L=20$ cm and for three different relative phases: (a) $\Delta \varphi =0 $, (b) $\Delta \varphi =\pi/2 $ and (c) $\Delta \varphi =\pi $. Inset in each figure shows the corresponding rotary photon drag vs angular velocity of host medium $\omega_{s}$. The plots are calculated for the probe detuning 0 $s^{-1}$. Other parameters are the same as in Fig.~\ref{figure2}(c). }
\label{figure5}
\end{figure} 

As concluded from previous results (Figs.~\ref{figure3} and \ref{figure4}), the relative phase can be used as a knob to change the group index as well as the nonlinear response (EIT, EIA and gain) and subsequently, we can efficiently control the photon drag by the relative phase. In the following, we concentrate on investigating transverse or lateral photon drag as the medium undergoes translational motion with a velocity changing from 0 m/s to $\pm 5$ m/s. As shown in Fig.~\ref{figure5}, the incident light is not displaced from its original path when the medium is stationary (as $ \Delta x=0$ for the case of $\mathrm{v}=0$), however, when it strikes the moving fast or slow-light medium, the light can be dragged. 

For $\Delta \varphi =0 $, light is dragged opposite to the motion, resulting in negative displacement at positive $\mathrm{v}$ and vice versa (see Fig.~\ref{figure5}(a)). For two other cases, the light is dragged along the medium translation as both $\Delta x$ and $\mathrm{v}$ have got identical signs. It is true that the value of lateral drag for a fixed velocity is not substantially enhanced as the light undergoes subluminal propagation for $\Delta \varphi =\pi $, compared to the superluminal one with $\Delta \varphi =0 $, but it is at least equally important to have a loss-less or gain-assisted light propagation. So, favorable results including an enhanced photon drag and a loss-less or gain-assisted light propagation are observed by changing the relative phase. Furthermore, the value of the lateral drag at a maximum translational velocity ($\pm $5 m/s) is enhanced by four times by tuning the relative phase from $\pi/2$ to $\pi$. At a more realistic value of the velocity of 1 m/s, the photon drag for $\Delta \varphi =\pi $ are about 0.03 m. It is worth pointing out that the lateral light drag takes place because of drag of light polarization state owing to change in group index according to Fig.~\ref{figure3}. 

Inset in Fig.~\ref{figure5} elucidates rotary photon drag for the same parameters of Fig.~\ref{figure2}(c). As shown in Fig.~\ref{figure5}(a), for $\Delta \varphi =0 $, light dragging takes place opposite the rotation of the optical medium; that is, if the medium rotates in clockwise direction, the dragging will be anticlockwise direction and vice versa. However, the rotary photon drag for $\Delta \varphi =\pi/2 $ and $\pi$ takes place along the rotation of the medium. By changing the phase from $\pi/2$ to $\pi$, the degree of the light polarization drag is enhanced from $0.16$ radians to $0.65$ rad for fixed angular frequencies, $\omega_{s}=\pm 20$ rps. One must, just like the case of the lateral photon drag, concentrate on the gain-induced photon drag appeared in Fig.~\ref{figure5}(c).  

Before ending this section, we note that our suggested scheme has the following key advantages: 1) The main advantage of our proposed scheme is its simple implementation as the relative phase could be easily changed by electro-optical devices. We must reiterate the importance of the fact the phase controlling of the phenomena, from experimental point of view, is much easier than that via the intensity or the frequency which worked out in previous studies \cite{sankar,khan,iqbal}. 2) What is more, our model shows a wide range of tunability so that a large gain can be achieved for a proper choice of the relative phase of applied fields. 3) The main feature of our suggested scheme, besides appearing a gain in the absorption spectra, is that both lateral and rotary photon drag are improved, compared with the similar works \cite{khan,iqbal} in which the maximum value of the rotary and lateral drag, respectively, was about $0.03$ radians at $\omega_s= 20$ rps and $10^{-3}$ m at $\mathrm{v}=$ 1 m/s. Moreover, calculated Fresnel drag coefficient, i.e., $F= n_{g}/n_{r}-1/n_{r}^{2}$ \cite{Landau} for the case of $\Delta \varphi =\pi$ would be about $5 \times 10^{7}$ where is also enhanced by at least two orders of magnitude, compared with previous experimental works \cite{safari,72}. Recalling that by considering the dispersion relation in the rest frame, employing the Lorentz transformation to the first order of $\mathrm{v}/c$, and then expanding the index of refraction in a power series of $k_{p} \mathrm{v}$ to the first order, the phase velocity $\mathrm{v}_p$ can be written as $\mathrm{v}_p=c/n_r + F \mathrm{v}$ \cite{72}, with the definition of the phase velocity $\mathrm{v}_p=\omega_p/k_p$ and the probe's wave-number $k_p$. 

The last part of the paper is devoted to measurement of the photon drag effect, by taking a glance at its potential applications. General speaking, in order to detect the light dragging effect, the phase of probe field must be compared with that of a local oscillator \cite{72,chen}. As the wave-number of the probe field in the moving medium is given by $[(\omega_p/c) (n_r+\mathrm{v} /c)] /(1+\mathrm{v}/\mathrm{v}_g)$, therefore, its phase shift passing through the medium can be written as 
\begin{equation}
\Psi =\dfrac{L \omega_p n_r}{c (1+ \mathrm{v}/\mathrm{v}_g )}.
\label{egn}
\end{equation}
Thus, by comparing such phase with the local oscillator, we are able to extract the phase shift of light as $(-k_p L\, n_{r} \mathrm{v})/\mathrm{v}_g$ for $\mathrm{v} \ll \mathrm{v}_g$. As can be seen from equation \ref{egn}, for the modest values for the probe field and for a minimum detectable phase \cite{sankar}, we can easily measure very small velocities. Using another method of detecting \cite{72}, moving the medium cell at some frequency as in Ref. \cite{modul}, when the satellites at the frequency of modulation appear near the probe frequency, the intensities of the satellites are proportional to $(\omega_p L \mathrm{v} /c \mathrm{v}_g)^{2}$, giving us the same sensitivity to the velocity. As the frequency of modulation used in the experiment \cite{modul} is about 100 kHz, so the detectable displacement is of the order of a few $ \angstrom$, much smaller than their used wavelength. Such scheme, showing the accuracy of the position of optical elements which can be controlled by detecting this signal, can be important for ultra-short physics and microscopy. 

%The enhanced dragging effect also proves efficient coupling between optical fields and mechanical motion, in particular sound waves.Other applications of the effect include its use for position control and the efficient modulators of light, as well as the detection of slow motion.  
  
\section{Conclusion}

In summary, we investigated the photon drag in a moving three-level $\Lambda$-type atomic system in which phase-dependent behavior is induced by the quantum interference due to the spontaneous emission. Optical properties of a weak probe field in such system can be dramatically modified by controlling the relative phase of applied fields, allowing us to further enhance the photon drag. Another prominent note, as far as the drag is concerned, is a loss-less or gain-assisted light propagation; We showed that an enhanced drag along with a gain can be achieved for a proper choice of the relative phase. It is worthy of mention that phase controlling of the photon drag is easier than that via the intensity or the frequency. Note that the photon drag is attributed to the dragging of light polarization state, owing to change in group index. Applications of such enhanced dragging effect include the position control, the efficient modulators of light and detection of slow motion.

\bibliographystyle{}

\end{document}